\documentclass[twocolumn,10pt,showpacs]{revtex4-1}
\usepackage[latin1]{inputenc}
\usepackage{amsmath}
\usepackage{amsfonts}
\usepackage{amssymb}
\usepackage{graphicx,hyperref}
\usepackage{amssymb}
\usepackage{mathptmx}
\usepackage{listings}
\newcommand{\be}{\begin {equation}}
\newcommand{\tcblu}{\textcolor{black}}

\newcommand{\ee}{\end {equation}}
\newcommand{\beqa}{\begin {eqnarray}}
\newcommand{\eeqa}{\end {eqnarray}}
\newcommand{\mb}{\mathbf}
\hypersetup{
    colorlinks,%
    citecolor=blue,%
    filecolor=blue,%
    linkcolor=blue,%
    urlcolor=blue
}

\begin{document}

\title{Parallel Implementation of 3D Molecular Dynamic Simulation for Laser-Cluster Interaction}

\author{Amol R. Holkundkar}
\email[E-mail: ]{amol.holkundkar@pilani.bits-pilani.ac.in}
\affiliation{Department of Physics, Birla Institute of Technology and Science, Pilani - 333 031, India.}

\begin{abstract}
The objective of this article is to report the parallel implementation of the 3D molecular dynamic simulation 
code for laser-cluster interactions. The benchmarking of the code has been done by comparing the simulation results 
with some of the experiments reported in the literature. Scaling laws for the computational time is established
by varying the number of processor cores and number of macroparticles used. The capabilities
of the code are highlighted by implementing various diagnostic tools. To study the dynamics of the laser-cluster 
interactions, the executable version of the code is available from the author. 
\end{abstract}


\maketitle

\section{Introduction}
The interaction of laser with atomic and molecular clusters has proved to be a challenging area of research 
(both experimentally as well as theoretically) over the last decade and it continues to do so.  The laser-cluster fraternity 
has constantly been working on new experimental and theoretical techniques to shed new light on the dynamics of 
the laser interaction with atomic and molecular clusters. The pioneer work in this field has been done by Ditmire et. al. \cite{ditmire1997_nature},
where they have shown the production of high energy electrons and protons by the explosion of the noble gas 
clusters.  However, the generation of intense x-rays reported by Parra et. al. \cite{parra2000_pre}, and the fusion 
reaction in Deuterium clusters have been reported by Zweiback et. al. \cite{zweiback2000_prl}. Also, Donnelly et. al. reported the generation 
of high order harmonics generation in laser-cluster interactions \cite{donnelly1996_prl}.

These various applications are possible as the clusters retain the advantages of an overall low-density gas 
and debris free operation and offer very high local solid density leading to strong 
laser heating. Hence, the dynamics of laser interaction with atomic clusters is completely different 
A number of theoretical studies on laser-cluster interactions have been conducted. The nanoplasma
model for the same is proposed by Ditmire et. al. \cite{ditmire1996_pra}. In this model, the cluster is considered 
to be a spherical plasma ball with uniform radial density and its evolution is modeled using the hydrodynamic 
approach. Milchberg et. al. \cite{milchberg2000_pre} have proposed a new model in which the non-uniformities 
in the hydrodynamic parameters are taken into account by achieving space variation of the electric field within the cluster
using near field approximation. Earlier we have also developed a 1-D Lagrangian hydrodynamics to understand the dynamics 
of laser-cluster interactions by treating the cluster as a stratified plasma ball so that the radial variation of the 
resulting plasma density can be modeled \cite{amol2008_pop}.

So far we have mentioned about the fluid approach to the laser-cluster interactions, but as clusters are just a collection of few
thousands of atoms or molecules bonded with weak Vander-Wall forces,  treating the problem under the purview 
of the hydrodynamic approach can be questioned sometimes. In view of this, particle models like particle-in-cell 
(PIC) and molecular dynamics (MD) are extensively used to study the dynamics of laser-cluster interactions 
\cite{taguchi2010_oe,petrov2005_pre}. The inherent advantage of MD over PIC is the absence of any computational grid which
makes it quite simple to implement in three dimensions. Undoubtedly,  PIC is a more rigorous method where in a whole set of Maxwell's equations is solved on a computational grid. On the other hand, MD is more suited for problems where in the number of particles is 
limited to a few thousands. 

The molecular dynamic approach to laser-cluster interactions was initiated by Last and Jortner \cite{last1997_jcp,last2000_pra,last2001_PRL,last2012_pop}. They have further shed light on the techniques which are very essential to 
perform the MD simulation in an efficient manner, like the lumping of actual particles to form a one macro-particle with same 
charge to mass ratio \cite{last2007_pra}.

A MD code is earlier developed by us  to study the dynamics of the laser interactions with the atomic clusters 
\cite{amol2011_pop}. But that code being a serial has its limitations in terms of the computational time required to simulate 
the dynamics of larger clusters or the interactions of longer pulses. To overcome this drawback,  a 
parallel version of the code is developed which could easily simulate the larger clusters. In this paper the said model is discussed along with its implementation followed by some bench marking results. 

\section{Simulation Model}

The dynamics of the laser-cluster interactions is modeled by treating the cluster as a sphere of radius $R_0$ placed at the center
of the coordinate system ($x = y = z = 0$). Dimensionless units are used in the model, so the electric field
is denoted by dimensionless units $a_0 = e E /m_e\omega c$, magnetic field by $b_0 = e B /m_e\omega$. Furthermore, time and space
are normalized with respect to $\omega$ and $k$ respectively. In the above parameters, $e$, $m_e$ are the electron charge 
and mass, $E$ and $B$ are the electric and magnetic fields in SI units, $\omega$ and $k$ are the laser frequency and 
wave vector, and $c$ is the speed of light. The number of atoms in a cluster is calculated by the knowledge of 
Wigner-Sietz radius $R_W$\cite{note}, 
\be N_{Atoms} = \left(\frac{R_0}{R_W}\right)^3 \label{no_atoms}.\ee 
The value of $R_W$ depends on the species of the gas cluster under study. For example, $R_W$ is 1.70, 2.02, 2.40 and 2.73 
\AA\ for Deuterium, Neon, Argon and Xenon clusters respectively \cite{petrov2008_pop}. 

Initially these atoms are randomly placed inside the cluster of radius $R_0$ (dimensionless). For computational purposes 
 predefined number of macro particles, $N_{Macro}$ are used. Each macro particle consists of $N_{Atoms}/N_{Macro}$ of 
actual particles (ions or electrons). This procedure of lumping of actual particles to form pseudo-particles is similar to the one 
presented by Last and Jortner \cite{last2007_pra} and Petrov et. al. \cite{petrov2005_pop}. The presence of other clusters in the neighborhood is simulated by imposing 
periodic boundary conditions on the faces of cubical simulation box with each side being considered equal to the inter cluster 
distance, $R_{IC} (= 20 R_0)$ \cite{petrov2008_pop}. However, the options for an open boundary or a combination of periodic 
and open boundary conditions are also incorporated in the code. \tcblu{The advantage of the mixed boundary condition is to ensure 
the relative faster ionization of cluster atoms as compared to the condition when only open boundary  is used. This is due to the fact that, 
the periodicity of the boundaries  mimics the presence of the neighboring clusters which in a sense can escalate the ionization by the 
collisional means. Then the open boundary condition ensures the free flight of the particles who no longer are affected by the neighboring clusters.}

\begin{figure}[t]
\centering \includegraphics[width=.7\columnwidth,angle=270]{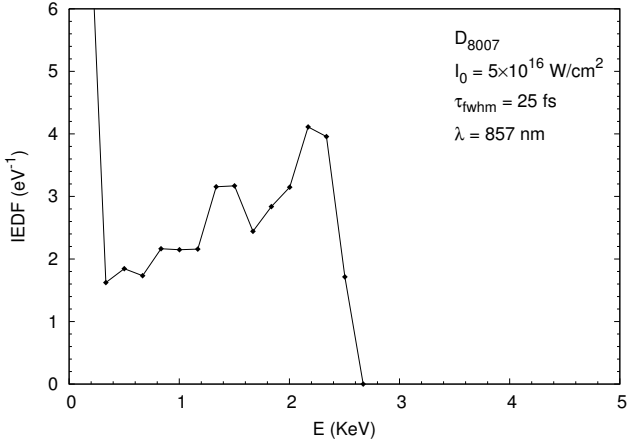}
\caption{Time integrated ion energy distribution  function (IEDF) for D$_{8007}$ clusters irradiated by 25 fs, 857 nm laser pulse
with peak intensity $5\times10^{16}$ W/cm$^2$. }
\label{iedf_deu}
\end{figure}

\begin{figure}[t]
\centering \includegraphics[width=.7\columnwidth,angle=270]{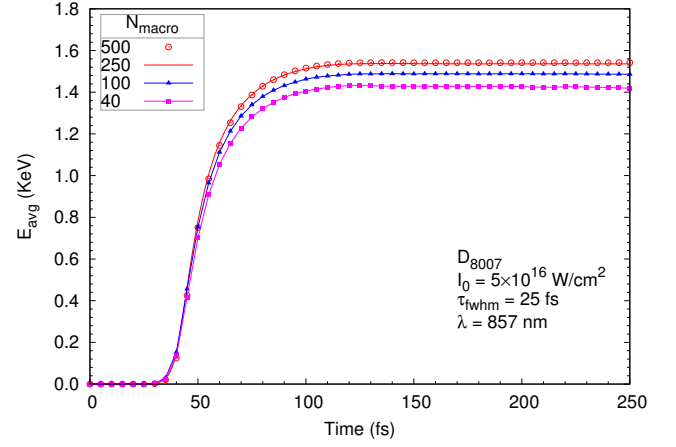}
\caption{\tcblu{Temporal evolution of the average energy of D$_{8007}$ cluster irradiated by 25 fs, 857 nm laser pulse
with peak intensity $5\times10^{16}$ W/cm$^2$. Each curve represents simulations of the cluster with a fixed number of macroparticles.}}
\label{macroDeu}
\end{figure}

\begin{figure}[b]
\centering \includegraphics[width=.7\columnwidth,angle=270]{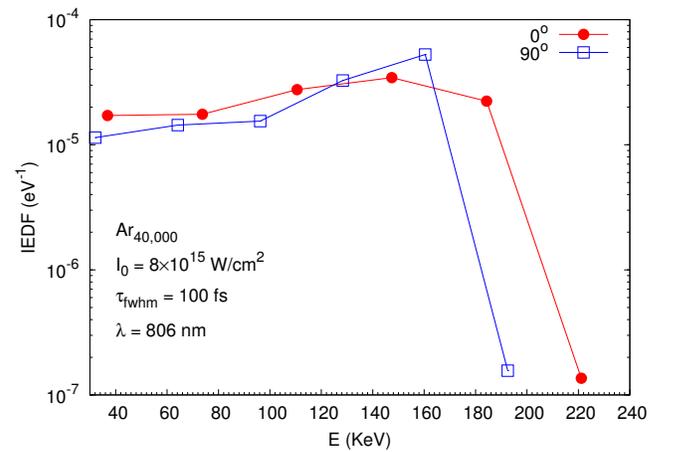}
\caption{Time integrated ion energy distribution function (IEDF) of Ar$_{40,000}$ measured along and perpendicular to laser
polarization direction.}
\label{iedf_aniso}
\end{figure}

\begin{figure}[t]
\centering \includegraphics[width=.7\columnwidth,angle=270]{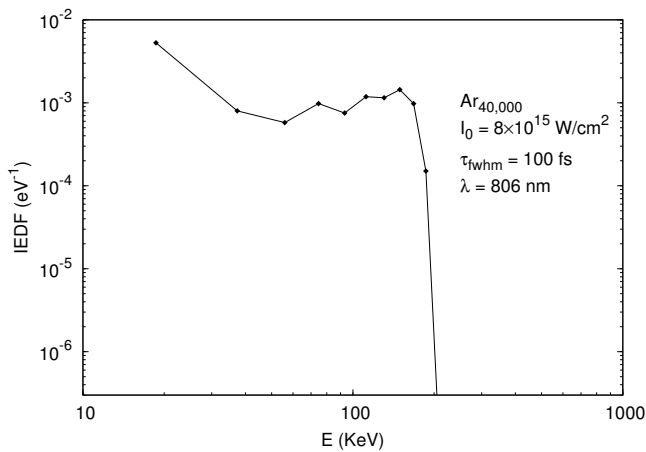}
\caption{Time integrated ion energy distribution (IEDF) function for Ar$_{40,000}$ clusters. }
\label{iedf_ar}
\end{figure}

\begin{figure}[b]
\centering \includegraphics[width=.7\columnwidth,angle=270]{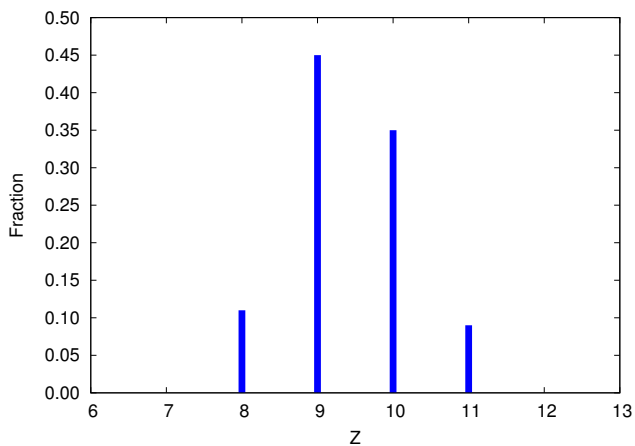}
\caption{Fraction of different Argon charge states at 500 fs. Laser and cluster parameters are same as in Fig. \ref{iedf_ar}.}
\label{z_snap}
\end{figure}

\begin{figure}[t]
\centering \includegraphics[width=.7\columnwidth,angle=270]{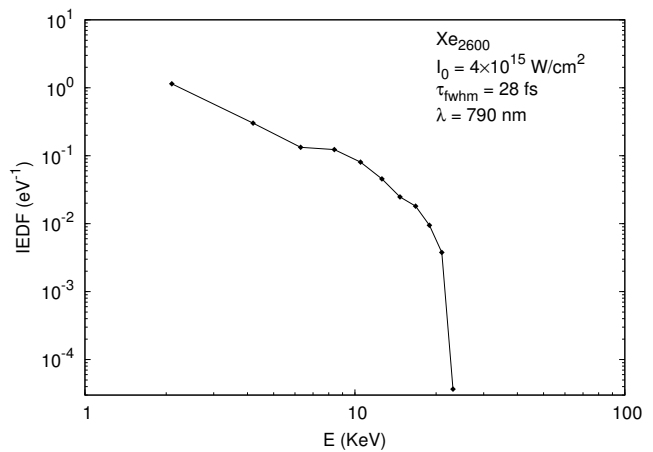}
\caption{Time integrated ion energy distribution function for Xe$_{2600}$ clusters irradiated by 28 fs, 790 nm laser pulse
with peak intensity $4\times10^{15}$ W/cm$^2$. }
\label{iedf_xe}
\end{figure}

The dimension of the cluster is very small compared to the laser wavelength, so only temporal variation in the
laser intensity is considered. The cluster is put in constant radiation bath of the electric field of the laser. The laser electric
field amplitude, in dimensionless units is calculated by the knowledge of the laser intensity as, 
\be a_0 = \sqrt{I_0 \lambda^2/1.366\times 10^{18}}.\ee Here $I_0$ is laser intensity in the units of W/cm$^2$ and $\lambda$ is
laser wavelength in units of $\mu m$. The temporal profile of the laser intensity is given by,
\be I(t) = a_0\ \exp[-2.77(t - 2 \tau_{fwhm})^2/\tau_{fwhm}^2].\ee Here $\tau_{fwhm}$ is the FWHM pulse length of the laser
($t$ and $\tau_{fwhm}$ should be taken in same units, whether SI or dimensionless units). The electric and magnetic
fields are given to be,
\beqa 
 E_x &=& a_0\ \sin(t) \\
 E_y &=& a_0\ \sin(t+\phi_E)\\
 E_z &=& 0\\
 B_x &=& b_0\ \sin(t+\phi_B) \\
 B_y &=& b_0\ \sin(t)\\
 B_z &=& 0.
\eeqa
 where values of $\phi_E$ and $\phi_B$ are taken as $3\pi/2$ and $\pi/2$ respectively for circularly polarized light. \tcblu{However 
 for linearly polarized light, its values are  chosen in such a way that $E_y$ and $B_x$ components of the laser field is zero.}
 Here $t$ is in dimensionless units (normalized to $\omega$). It should be noted that the electric field corresponding 
 to $a_0 = 1$  signifies the corresponding magnetic field to be $b_0 = 1$. With these criteria, it is convenient to represent 
 both the electric and magnetic field using same parameters. 

The ionization of the cluster atoms is carried out by using optical field and collisional ionization through Monte-Carlo method. 
For the optical field ionization (OFI), we calculate the OFI rate $\nu_{ofi}$ as a function of laser field strength using 
ADK tunnel ionization formula \cite{ADK1986_zphy} \citep{amol2011_pop}. The collisional ionization rate $\nu_{ci}$ is 
calculated from the best fitted data on the ionization rates \cite{voronov1997_adt} for atomic species of the cluster 
whose $Z < 28$. However, Lotz formula for the collisional ionization rate is used for Xe, which  is calculated as mentioned in Ref. \cite{lotz,amol2011_pop} for a particular ionic species. Once total ionization rate, $\nu_{total} = \nu_{ofi} + \nu_{ci}$, for an atom
is calculated, a random number, $0 < \alpha \leq 1$, is generated and compared with the 
 $P = 1 - \exp[-\nu_{total} \Delta t]$, where $\Delta t$ is a time step for the calculation of the ionization process. If 
$\alpha < P$ then the atom is ionized and an electron is created randomly in the vicinity of the atom \citep{amol2011_pop}.  

The particles are advanced under the influence of Lorentz and Coulomb forces due to the presence of other charged particles.
Relativistic effects are taken into consideration while formulating the equation of motion. The standard Boris leap frog 
algorithm is used where in the motion in electric and magnetic fields is decomposed and $\mb{v} \times \mb{B}$ rotation is 
properly achieved \cite{filippychev}. Boris algorithm is  numerically stable  and it is not very sensitive to the choice 
of time step. 

The \textsc{OpenMP} framework is used to execute the program in parallel. It harnesses the multi-core architecture of 
any modern processor. As the particle mover and force calculation consumes lot of computational time,  only these two modules
are paralleled using \textsc{OpenMP}. The program is executable on any modern Linux desktop or server with GNU compiler g++ version
greater than 4.0.  

\section{Results and Discussion}

In order to validate the simulation model, we simulated some experiments and existing simulation results for Argon, Xenon and Deuterium clusters. It is observed that the energy spectrum of energetic ions in all experiments is in good agreement with the results obtained from the MD code, which indirectly signifies that the dynamics is correctly simulated. 

To begin with   a case with D$_{8007}$ cluster interaction with 25 fs, 
857 nm laser with peak intensity of $5\times10^{16}$ W/cm$^2$ is simulated using 500 macroparticles. The time 
integrated ion energy distribution function for this case is presented in Fig. \ref{iedf_deu}, which shows 
good agreement with the results presented by \cite{last2001_PRL}. As expected the average kinetic energy is found to
be $\sim$ 1.5 keV, which in this case is 3/5$^{th}$ of the maximum energy observed ($\sim$ 2.5 keV). 

\tcblu{
To validate the effectiveness of the lumping procedure, the same case as in Fig. \ref{iedf_deu} is simulated with number of 
marcoparticles ranging from 40 to 500. The time evolution of the average kinetic energy as a result of varying the number of 
macroparticles is presented in Fig. \ref{macroDeu}. It is clear from this figure that the 92\% change of the ratio 
$N_{Atoms}/N_{Macro}$ is reflected in just 7\% change in the final average energy of the D$_{8007}$ cluster. 
This indeed, establishes the stability of the lumping procedure in the simulations. Keeping the results from Fig. \ref{iedf_deu} as reference, 200 macroparticles are further used to simulate the  Ar$_{40,000}$ and Xe$_{2600}$ clusters. 
}

The anisotropic explosion of the Argon clusters was reported by Kumarappan et. al. \cite{kumar2001_PRL}. They had shown that the explosion of Ar$_{40,000}$ cluster after the interaction of the 100 fs, 806 nm laser pulse with peak intensity of $8\times10^{15}$ W/cm$^2$, results in an anisotropic ion energy distribution. It was observed that more energetic ions were detected along the laser polarization direction compared to those detected in the perpendicular direction. To understand the dynamics of this anisotropic cluster explosion,  the same case has been simulated with this MD code. The cluster radius is calculated using Eq. (\ref{no_atoms}) with $R_W = 2.4$ \AA, which is about 82 \AA. The detector is placed at a location 9 R$_0$ from the center of the simulation domain. Two detectors are placed along (x axis) laser polarization and two are placed perpendicular to it (y axis). The particles which are making an angle less than 30$^o$ with the detector normal are counted. The time integrated ion energy distribution along (0$^o$) and perpendicular (90$^o$) to laser polarization is presented in Fig. \ref{iedf_aniso}. It can be seen that more energetic ions are emitted along the laser polarization direction compared to those in the direction perpendicular to it.
Complete time integrated ion energy distribution is presented in  Fig. \ref{iedf_ar}.  Energies which are seen in Fig. \ref{iedf_aniso} and \ref{iedf_ar} ($\sim$ 200 keV) are in good agreement with the experimental findings. The anisotropic
explosion of the cluster is explained on the basis of charge-flipping model \cite{kumar2001_PRL}. Rapid oscillations of the 
electrons in the laser field cause the atoms on the laser polarization axis to be more ionized resulting in more 
energetic ions along the laser polarization compared to their number in direction perpendicular to it. The fraction of the charge states at 500 fs
is presented in Fig. \ref{z_snap}, the observed charge states are also in good agreement with the experimental findings \cite{kumar2001_PRL}.

\begin{figure}[t]
\centering \includegraphics[width=3.5in,height=3in]{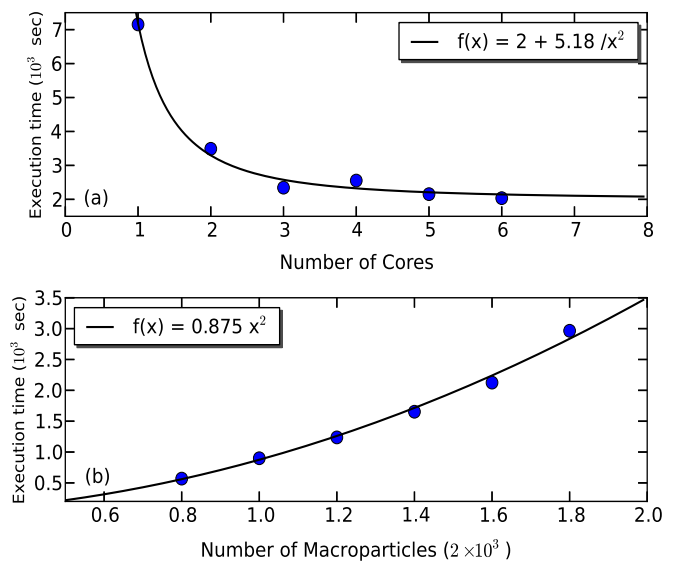}
\caption{CPU execution time with increasing number of cores and number of macroparticles. \tcblu{The CPU is \textit{Intel(R) Xeon(R) CPU E5645  @ 2.40GHz}. }}
\label{time_core}
\end{figure}

The experiments of cluster interactions with comparatively shorter laser pulses ($\sim$ 30 fs) are conducted by Skopalov\`{a} et al. \cite{skopalova2010_PRL}. They have studied the interaction of Xe$_{2600}$ cluster with
28 fs, 790 nm laser pulse with peak intensity of $4\times10^{15}$ W/cm$^2$. They showed the anisotropic explosion of 
the Xenon clusters, but not the one   mentioned in the last paragraph. They observed that more energetic ions were detected  
in the direction perpendicular to the laser polarization and not along it. This kind of anisotropy is attempted to be simulated but were  the phenomenon mentioned above is not observed. However, the energies of the Xenon ions   observed in the simulations are in good agreement
with their experimental findings. Time integrated ion energy distribution for this case is presented in Fig. \ref{iedf_xe}, which seems to be in good agreement with the experiment performed by Skopalov\`{a} et al. \cite{skopalova2010_PRL}.

Finally we would like to shed some light on the performance of the parallel MD code using \textsc{OpenMP} framework.
The shared memory architecture of the \textsc{OpenMP} framework makes it attractive to write a parallel code
compared to an MPI framework. Now a days CPUs are equipped with multi-core processors, typically 12 or more cores.
In view of this we decided to harness the power of multi-core processors by using all the cores of the same CPU. 
It is done by avoiding the use of MPI framework, which involves the communication with other CPUs in high speed network. 

To study the performance of the code,   the 80 \AA\ Deuterium cluster dynamics 
for 5000 time steps is simulated on \textit{Intel(R) Xeon(R) CPU E5645  @ 2.40GHz}. It is assumed that cluster is fully ionized at $t=0$ itself, and hence the code has to run by keeping equal number of electron and ion macroparticles. The execution time of the 
program with 3000 macroparticles (electrons + ions), executed with different number of cores, is presented in 
Fig. \ref{time_core} (a). The dots denote the actual time taken in seconds and solid line shows the power 
law fit to the curve. It is observed that the computation time scales as $\sim N_{core}^{-2}$, where $N_{core}$
is the number of cores used to run the code. The computational time with different number of total macroparticles
using 6 cores is shown in  Fig. \ref{time_core} (b). The different number of macroparticles  effectively changes the lumping
parameter of 80 \AA\ Deuterium cluster. It is observed that computational time increases as $\sim N_{part}^{2}$, where
$N_{part}$ is the number of total macroparticles used in simulation.

\section{Conclusion}
A parallel implementation of the molecular dynamic simulation for laser-cluster interactions is presented. The results
from the MD simulations are compared to the existing experiments and simulations. The ion energy distributions of 
Ar$_{40,000}$, Xe$_{2600}$ and D$_{8007}$ clusters are found to be in agreement with the existing results. The anisotropic
explosion of Ar$_{40,000}$ cluster with energetic ions along the laser polarization direction is also observed in the
simulations. The scaling laws for the computational time with number of processor cores and number of macroparticles 
are also established. It is found that computation time scales as $\sim N_{core}^{-2}$ and  $\sim N_{part}^{2}$
with number of cores and macroparticles are used. The stability of the code is also been established by varying the lumping parameter
without affecting the simulation results adversely. 

The executable version of the code can be obtained from the author.

\section{Acknowledgments}
The author is thankful to G. M. Petrov for various useful discussions on the dynamics of laser-cluster interaction. Author is 
also thankful to Kumar S. Bhattacharya for proofreading the manuscript.


\begin{thebibliography}{99}
\bibitem{note} Number of atoms are anyway dimensionless, so here $R_W$ we can consider in \AA\ and so the cluster radius.
\end{thebibliography}
\end{document}